\documentclass{emulateapj}
\shorttitle{Spitzer Observations of APM08279+5255}
\shortauthors{Soifer  et al.}
\slugcomment{ApJSup accepted (Spitzer Special Issue)}

\begin{document}

\title{Spitzer/IRS\altaffilmark{1}\ Observations of the Redshift 3.91 quasar APM 08279+5255\altaffilmark{2}}

\author{B.T. Soifer\altaffilmark{3}, V. Charmandaris\altaffilmark{4,5}, 
B.R. Brandl\altaffilmark{4,6}, 
L.Armus\altaffilmark{3}, P.M. Appleton\altaffilmark{3}, M.J. Burgdorf\altaffilmark{3}, 
D. Devost\altaffilmark{4}, T. Herter\altaffilmark{4}, S.J.U. Higdon\altaffilmark{4}, 
J.L. Higdon\altaffilmark{4}, J.R. Houck\altaffilmark{4}, C.R. Lawrence\altaffilmark{7}, 
P.W. Morris\altaffilmark{3}, 
H.I. Teplitz\altaffilmark{3}, K.I. Uchida\altaffilmark{4}, J. van Cleve\altaffilmark{4,8}, 
D.Weedman\altaffilmark{4}}

\email{bts@ipac.caltech.edu}

\altaffiltext{1}{The IRS was a collaborative venture between Cornell
University and Ball Aerospace Corporation funded by NASA through JPL
and the Ames Research Center.}  \altaffiltext{2} {based on
observations obtained with the Spitzer Space Telescope, which is
operated by JPL, California Institute of Technology for the National
Aeronatics and Space Administration}

\altaffiltext{3} {Spitzer Science Center, California Institute of
Technology, 314-6, Pasadena, CA 91125}

\altaffiltext{4} { Cornell University,  Astronomy Department,
Ithaca, NY 14853-6801}
\altaffiltext{5} { Chercheur Associ\'e, Observatoire de Paris, F-75014, Paris,  France}
\altaffiltext{6} { Leiden University, 2300 RA Leiden, the Netherlands}
\altaffiltext{7} {Jet Propulsion Laboratory, California Institute of
Technology, Pasadena, CA 91109}
\altaffiltext{8} {Ball Aerospace Technologies Corp., 1600 Commerce St., Boulder, CO 803001}

\vfill\eject

\begin{abstract}

The Infrared Spectrograph (IRS) onboard the Spitzer Space Telescope
(SST) has been used to obtain low and moderate resolution spectra of
the dust and gas-rich quasar APM~08279+5255 (z=3.91).  Broad
Paschen~$\alpha$ and $\beta$ recombination lines of hydrogen were
detected at wavelengths of 9.235 and 6.315$\mu$m, as well as a strong,
red continuum that is a smooth power law over the observed (rest
frame) wavelength range 5.3-35$\mu$m (1.08 - 7.1$\mu$m).  The observed
P$\alpha$/P$\beta$ line flux ratio of 1.05$\pm$0.2 is far from the
case B value of $\sim$2 and simple models of high density, high
optical depth ionized gas regions ($\sim$1.8). This deviation is
opposite in sense to the expected effect of reddening.

No evidence is found in the spectrum for either the 3.3$\mu$m or
6.2$\mu$m emission features usually attributed to aromatic
hydrocarbons in gas rich galaxies in the local Universe.  This is
consistent with the high luminosity AGN nature of APM~08279+5255.

\end{abstract}


\newpage

\section{Introduction}

The quasar APM~08279+5255 was originally discovered in a survey for
high galactic latitude carbon stars(Irwin et al. 1998).  At a redshift
z=3.91 (Downs et al. 1999), its brightness makes APM~08279+5255
apparently the most luminous object known in the universe. As
originally suggested by Irwin et al.  and subsequently confirmed by
Ibata, et al.  (1999) and Egami et al. (2000) APM~08279+5255 is
strongly gravitationally lensed, with a magnification of
$\sim$100. The lensing reduces its intrinsic bolometric luminosity to
a more modest but still large $\sim$5$\times 10^{13}$L$_{\odot}$.

Although APM~08279+5255 does not have a strong infrared excess for
quasars (see, e.g. Wilkes, 2001), it is a bright source in the IRAS
point source catalog (JISWG, 1989, Irwin et al. 1998).  It has been
detected in the millimeter and submillimeter continuum (Lewis et
al. 1998) and in multiple CO emission lines (e.g. Downs, et al. 1999,
Papadopoulos et al. 2001), demonstrating that the lensed quasar system
is dust and gas-rich.  The mass of molecular gas is 1 -- 10 $\times
10^9 M_{\odot}$, depending on the lensing magnification (Downs et al.,
Lewis, et al. 2002).  The presence of large quantities of gas and dust
in such a high redshift system makes it a prime candidate to search
for complex molecules at high redshift. The spectral energy
distribution of APM~08279+5255 at far infrared and submillimeter
wavelengths is well-fitted with a black body of temperature $\sim$220K
(Lewis et al.) and has a continuum level well above 70 mJy for
observed wavelengths $\lambda >$ 12$\mu$m.  The spectral energy
distribution of APM~08279+5255 is normal for radio quiet quasars. Its
unusual brightness makes it ideal for study by the Spitzer Space
Telescope (SST) as an example of a high redshift quasar, and makes it
of great interest for probing the nature of quasars at very large
lookback times.

APM~08279+5255 was therefore selected as an early release observation
target for the Infrared Spectrometer (IRS) on the SST.  The following
sections describe the observations and data reduction, the IRS data,
and their implications.

\section {Observations and Data Reduction}

APM08279+5255 was observed by the Spitzer/IRS on 26 October 2003 and
23 November 2003. In order to ensure proper target placement on the
IRS slits, the spectroscopic observations were preceded by target
acquisition in the IRS peak-up imager (Houck et al. this issue,
hereafter Paper I).  Table 1 presents the log of the Spitzer
observations.  For the low resolution spectra, observations were
obtained at two positions along the IRS slit to enable sky
subtraction.  High resolution spectra were obtained of both the target
and a blank nearby patch of sky that was used for sky subtraction.
Because of the substantially higher signal-to-noise ratios of the 23
November observations, where there was overlap between observations on
the two dates, we include only those data from 23 November in the
subsequent discussion.

\begin{deluxetable}{cccc}
\tablewidth{0pt}
\tablenum{1}
\tablecaption{Observation Log  }
\tablehead{
\colhead{$\lambda_{range}$[$\mu$m]}    &
\colhead{$\lambda$/$\Delta\lambda$}    &
\colhead{t$_{int}$[sec]}               &   
\colhead{Observation Date }                        
}
\startdata
10 -- 19.5 & 600 & 240 & 26 Oct\tablenotemark{a} \\
19 -- 37 & 600 & 240 & 26 Oct \\
7.5 -- 14 & 80 & 56 & 26 Oct \\
19 -- 37 & 600 & 480 & 23 Nov\tablenotemark{b}\\
5.3 -- 8.5 & 80 & 56 & 23 Nov \\
7.5 -- 14 & 80 & 56 & 23 Nov \\
14 -- 21.8 & 80 & 84 & 23 Nov \\
20.6 -- 40 & 80 & 84 & 23 Nov \\
\enddata
\singlespace
\tablenotetext{a}{600 second sky dark obtained on 23 Nov}
\tablenotetext{b}{300 second sky dark obtained on 23 Nov}
\end{deluxetable}

The data were processed through the IRS data pipelines at the Spitzer
Science Center (Paper I) to produce calibrated data frames.  The data
were converted from slopes to integrated signals, linearized, dark
subtracted and cosmic rays were removed. Flat field corrections were
applied, as were wavelength calibrations. Multiple frames covering the
same wavelength range were combined, and one-dimensional
sky-subtracted spectra of the target quasar were extracted from the
combined data.

The photometric calibration of the spectrum was based on Spitzer
observations of standard stars (Paper I), while the wavelength
calibration was based on observations of emission line stars and
planetary nebulae (Paper I).  At this time we estimate that the
overall uncertainty in the wavelength calibration is $\pm$0.04$\mu$m
at 20$\mu$m for the low resolution spectra, and $\pm$0.006$\mu$m at
20$\mu$m for the high resolution spectra.

\section {Results}

The low resolution spectrum, plotted as flux density
vs. observed(rest) wavelength, is plotted from 5--38 (1.1--7.7) $\mu$m
in Figure 1. No features were seen in the high resolution spectrum
that were not present in the low resolution spectrum.  Because of the
present uncertainties in the IRS calibration, the long-low spectrum
was scaled by factor of 0.78, to match with the short-low spectrum and
to bring the flux density into agreement with the flux density
observed at 18$\mu$m by Egami et al. (2000). The overall level agrees
well with the flux density reported at 12.5 $\mu$m by Egami et al.
With this adjustment the spectrum of APM08279+525 is nearly
featureless, being dominated by a remarkably smooth continuum.
Fitting the continuum by eye as a power law, the slope is
f$_{\nu}\sim\nu^{-1.21}$ for $\lambda >$ 15$\mu$m ($>$2.85$\mu$m in
the rest frame).  This continuum fits the IRAS measured flux densities
at 60 and 100 $\mu$m remarkably well. For wavelengths $\lambda <$
15$\mu$m ($<$2.85$\mu$m in the rest frame), the slope steepens
considerably, to f$_{\nu}\sim\nu^{-2.26}$.  There are no broad
emission features present in the spectrum.

\begin{figure*}
\figurenum{1}
\epsscale{1.2}
\plotone{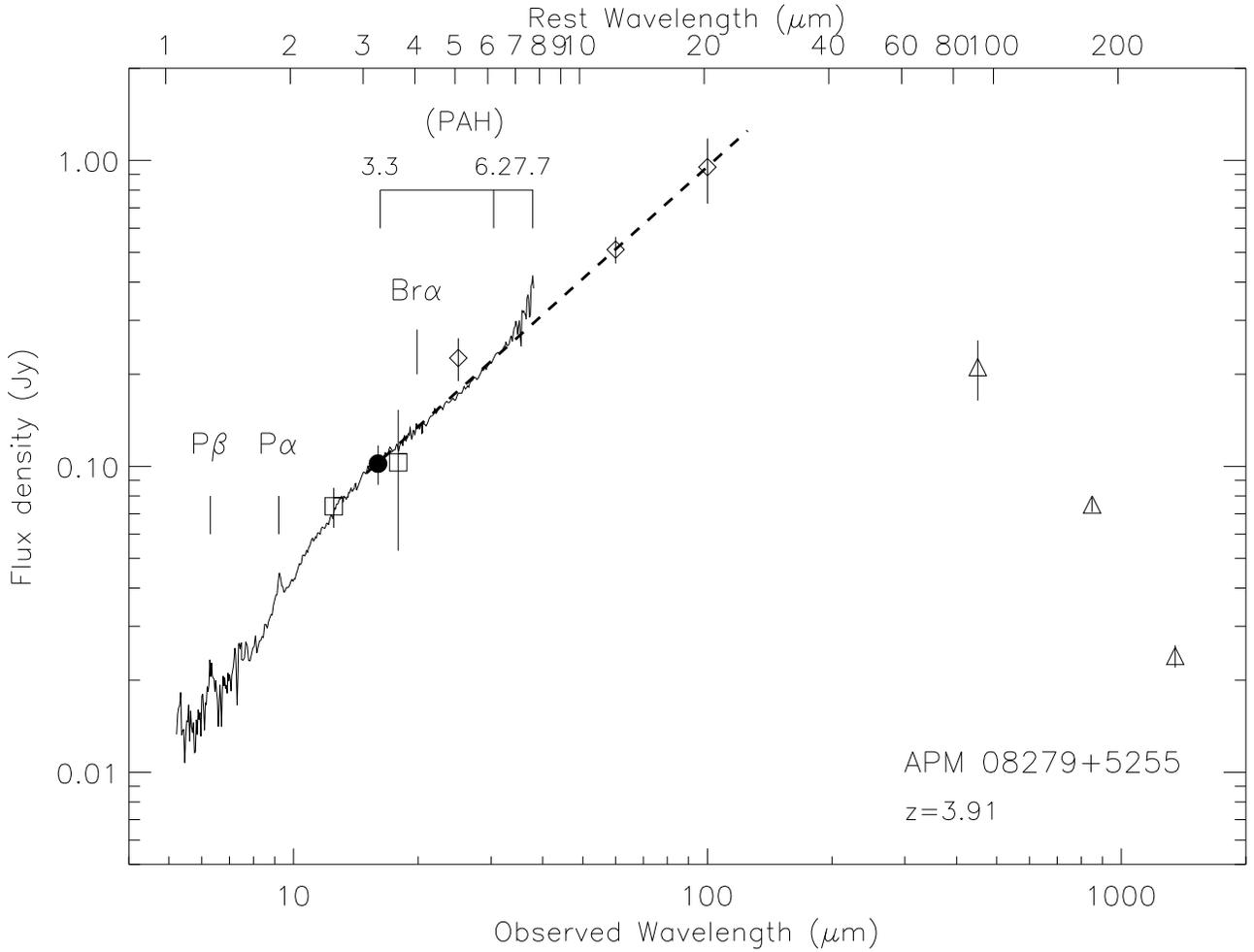}
                                                                                                   \caption{The mid-infrared spectra of APM08279+5255, plotted as flux
density vs. wavelength (observed - bottom, restframe - top). In
addition to the IRS spectrum, the flux density as measured in the IRS
peak-up camera at 16$\mu$m is plotted as the filled circle. The IRAS
flux densities at 25, 60 and 100 $\mu$m (JISWG, 1989) are plotted as
open diamonds, the flux densities measured by Egami et al. (2000) at
12.5$\mu$m and 18 $\mu$m are plotted as open squares, and the
submillimeter observations of Lewis et al. (1998) are plotted as open
triangles . A power-law fit to the continuum for
$\lambda_{observed}>$15$\mu$m is shown as a dashed line. Wavelengths
of hydrogen recombination lines and expected PAH features are
indicated in the plot.}
                                                                                                     
\end{figure*}

Since the background quasar is gas/dust rich, it might have been
expected that the emission bands of dust or organic molecules (PAH)
would be present, but none has been detected in the spectrum. Upper
limits to the fluxes of the PAH bands at 3.3 and 6.2$\mu$m are given
in Table 2.  The rise in the spectrum at the longest wavelengths
($>$35$\mu$m) might indicate the presence of a weak PAH feature at a
rest wavelength of 7.7$\mu$m, but the signal-to-noise ratio in this is
low, and flat fielding and calibration issues at the long wavelength
end of the spectrum make the reality of this feature questionable at
best. We therefore do not consider it real.

\begin{deluxetable*}{lcccccc}
\tablewidth{0pt}
\tablenum{2}
\tablecaption{Observed Line Properties in APM08279+5255}
\tablehead{
\colhead{Line}                             &
\colhead{Observed Wavelength}              &
\colhead{z}                                &   
\multicolumn{2}{c}{Equivalent Width}       & 
\colhead{$\Delta$v }                       &
\colhead{Observed Flux}                    \\
\colhead{}                                 &
\colhead{[$\mu$m]}                         &
\colhead{}                                 & 
\colhead{[observed $\mu$m ]}               &
\colhead{[rest $\mu$m]}                    &  
\colhead{ FWHM [km\,s$^{-1}$]}                     &         
\colhead{[W\,cm$^{-2}$]}                         \\
}
\startdata
Paschen $\alpha$ & 9.235$\pm$0.011 & 3.925$\pm$0.006 &0.059 & 0.012 &  8770 & 8.5$\pm$1.1$\times$10$^{-21}$\\
Paschen $\beta$  & 6.315$\pm$0.013 & 3.914$\pm$0.011 &0.061 & 0.012 &  9500 & 8.1$\pm$1.6$\times$10$^{-21}$\\
Brackett $\alpha$ &19.89 \tablenotemark{a}  & -- & $<$0.016 \tablenotemark{b} & 0.003 \tablenotemark{b} & -- & 
$<$1.6$\times$10$^{-21}$\tablenotemark{b} \\
3.3$\mu$m &  16.203 \tablenotemark{a}  & -- & $<$0.01\tablenotemark{b} & 0.002 \tablenotemark{b} & -- & 
$<$1.0$\times$10$^{-21}$\tablenotemark{b} \\
6.2$\mu$m &  30.442 \tablenotemark{a}  & -- & $<$0.03\tablenotemark{b} & 0.006 \tablenotemark{b} & -- & 
$<$2.3$\times$10$^{-21}$\tablenotemark{b} \\
\enddata
\singlespace
\tablenotetext{a}{assumed redshift 3.91}
\tablenotetext{b}{3$\sigma$ upper limit}

\end{deluxetable*}

The only spectral features detected are emission lines at wavelengths
of 6.315$\mu$m and 9.235$\mu$m.  These features correspond to
redshifted hydrogen recombination lines Paschen $\beta$ and Paschen
$\alpha$, respectively, in the spectrum of the lensed quasar. The
lines are well resolved even in the low resolution spectra presented
here (R=$\Delta\lambda$/$\lambda\sim$80, Paper I).  The central
wavelengths, widths, and fluxes of the lines are presented in Table 2,
as are the upper limits of the line flux and equivalent width for
Brackett $\alpha$.

\section {Discussion}

The most notable features of the infrared spectrum of APM08279+5255
are the smooth, steep continuum, and the Paschen recombination lines.
The continuum that rises smoothly to the longest observed wavelengths
is consistent with a model of dust in an accretion disk heated by an
underlying central source.  Such an interpretation of the infrared
continuum of quasars has been presented by Phinney (1989) and Sanders
et al.(1989).  These models have been extended to toroidal disks by
Pier and Krolik (1992, 1993) and clumpy media by Elitzur, Nenkova and
Ivezic (2003).

The presence of substantial emission at rest wavelengths
$\lambda_{rest}>$2$\mu$m argues for temperatures approaching the
sublimation temperature of refractory dust.  The change in the
continuum slope at a rest wavelength of $~$3$\mu$m is consistent with
a decrease in dust abundance for T$_{dust} >$1000K, consistent with
the sublimation temperature of silicate grains (Salpeter, 1974).  The
overall continuum of APM~08279+5255, as mentioned above, is rather
typical for radio quiet quasars, without a pronounced infrared
peak. This suggests, in the "unified scheme" that we are viewing the
AGN accretion disk close to face on (c.f. Elitzur et al. 2003).

The non-detection of PAH features at 3.3$\mu$m and 6.2$\mu$m in the
spectrum of APM08279+5255 is consistent with the trend found in nearby
AGNs for Seyfert 1 systems and quasars to have weak or non-existent
emission features associated with PAHs (Tran et al. 2001). It should
be noted that evidence for PAH bands has been found in some
"infrared-loud" quasar spectra (e.g. Mk 1014 - see Armus et al. 2004),
so this is not a universal phenomenon.  Organic molecules such as PAHs
appear to be destroyed in the harsh environments of quasars. It is
likely that this process is at work in APM~08279+5255, so no
conclusions can be reached based on these observations about the
abundances of such molecules in the host galaxy of this quasar.

Another significant caveat regarding the interpretation of the
non-detection of PAH bands is the gravitational lensing, with the
potential for substantial differential magnification of different
regions within the source plane. This effect is clearly illustrated in
the three image model of Egami et al.(2000). For example a region of
hot dust emission located close to the caustics would be much more
magnified than a more extended source such as would be expected of a
PAH emitting region. This would have the effect of substantially
reducing the equivalent width of any PAH bands in the {\it observed}
spectrum.

The measured redshifts of the P$\alpha$ and P$\beta$ lines (Table 2)
are consistent with the redshift determined from the CO lines (Downs
et al. 1999), and greater than the redshift determined from the rest
UV spectral features (Irwin et al. 1998).  The line fluxes for the
P$\alpha$ and P$\beta$ lines (Table 2) lead to an observed line ratio
of P$\alpha$/P$\beta$=1.05$\pm$0.2. This ratio is substantially less
than predicted value of 2 from Case B recombination (Hummer and Storey
1987) and models that allow for high density and large optical depths
in the Paschen lines (Drake and Ulrich, 1980, Lynch et al. 2000),
which predict values for this ratio of 1.8-2. Reddening would drive
the line ratio to greater values, and so cannot be invoked to explain
the observations.  The upper limit on the flux in the Br $\alpha$
line, reported in Table 2, leads to a limit on the ratio of
Br$\alpha$/P$\alpha <$ 0.19.  This value is close to the case B value
of 0.24 and the high density, optically thick models of Drake and
Ulrich and Lynch et al. of $\sim$0.2. The limit on the ratio
Br$\alpha$/P$\beta <$0.2 is significantly smaller than the case B
ratio of 0.5, but is close to the Drake and Ulrich and Lynch et al.
models of $\sim$0.2 for this ratio. Clearly, more detailed physical
models of the broad line regions in the Paschen lines are needed to
account for the observed line ratios.

The ratio of the observed Paschen lines to Lyman~$\alpha$ and / or
H$\alpha$ would be most interesting to establish the physical
conditions in the broad line region and the reddening to this region.
Unfortunately the line profile of Lyman~$\alpha$ is strongly affected
by the intervening column of neutral hydrogen (Irwin, et al. 1998),
making straightforward comparisons fruitless.  In addition, H$\alpha$
falls longward of the wavelengths reachable with sensitive near
infrared spectrometers. H$\beta$ is potentially reachable from
groundbased telescopes, but no such observations have been published
to date.

The line widths from the fits to the Paschen lines ($~$9000
km\,s$^{-1}$) are consistent with those found for broad lines in
quasars.  Again, the observed broad UV lines in APM08279+5255 are
heavily affected by intervening absorption (e.g. Irwin, et al. 1998,
Hines, Schmidt and Smith, 1999), so no meaningful comparison of the
Paschen and UV linewidths can be made.

\section {Summary}

The observed mid-infrared spectrum of the high redshift lensed quasar
APM08279+5255 has shown recombination lines of P$\alpha$ and P$\beta$,
and a remarkably smooth overall continuum that rises sharply into the
far infrared.  The continuum shows a steepening in the continnum for
wavelengths $\lambda <$15$\mu$m (2.85$\mu$m in the rest frame of the
lensed quasar). This is suggestive of the temperature of sublimation
of silicate grains in a quasar accretion disk.  No features
characteristic of organic molecules, such as aromatic features, are
seen in the spectrum. The ratio P$\alpha$/P$\beta$ is far from that
predicted by models of the high density, high optical depth gas
believed to be in the quarar broad line region, in the sense opposite
to that expected from dust reddening.

\section {Acknowledgments}

We thank R. Puetter for helpful discussions.  It is a pleasure to
thank the many hundreds of people whose tireless efforts have lead to
the success of the Spitzer Space Telescope. In particular we
acknowledge the efforts of D. Gallagher, W. Irace and L. Simmons of
JPL, K.  Duclos of Cornell University and J. Marriot and J.
Troeltzsch of Ball Aerospace Technologies Corporation.

This work is based on observations made with the Spitzer Space
Telescope, which is operated by the Jet Propulsion Laboratory,
California Institute of Technology under NASA contract 1407. Support
for this work was provided by NASA through contract number 1257184
issued by JPL/Caltech.



\begin{thebibliography}{}

\bibitem {Armus04} Armus, L. et al. 2004, \apjs, (this volume)

\bibitem {Downs99} Downs, D., Neri, R., Wiklind, T., Wilner, D.J. and 
Shaver, P.A. 1999, \apjl, 513, L1

\bibitem {Drake80} Drake, S.A. and Ulrich, R.K. 1980,\apjs, 42, 351

\bibitem {Egami00} Egami, E., Neugebauer, G., Soifer, B.T., Matthews, K., Ressler, M., Becklin, E.E.,
Murphy, T.W. and Dale, D.A. 2000, \apj, 535, 561

\bibitem {Elitzur03} Elitzur, M., Nenkova, M, and Ivezic, Z. 2003, in {\bf The Neutral
ISM in Starburst Galaxies}, Aalto, S. and Huttemeister, S. eds., 
ASP Conference Series, astro-ph 0309040


\bibitem {Hines99} Hines, D.C., Schmidt, G.D. and Smith, P.S. 1999, \apj, 514, 91

\bibitem {Houck04} Houck, J.R., et al.2004, \apjs, (this volume, Paper I)

\bibitem {Hummer87} Hummer, D.G. and Storey, P.J. 1987, \mnras, 224, 801

\bibitem {Ibata99} Ibata, R.A., Lewis, G.F., Irwin, M.J., Lehar, J. and Totten, E.J. 
1999, \aj, 118, 1922   
 
\bibitem {Irwin98} Irwin, M.J.,  Ibata, R.A., Lewis, G.F. and Totten, E.J. 
1998, \apj, 505, 529

\bibitem {Lewis98} Lewis, G.F., Chapman, S.C., Ibata, R.A., Irwin, M.J. and 
Totten, E.J. 1998, \apjl, 505,L1

\bibitem {Lewis02} Lewis, G.F., Carilli, C., Papadopoulos, P. and Ivison, R.J. 2002,
\mnras, 330, L15

\bibitem {JISWG89}Joint IRAS Science Team, 1989, IRAS Point Source
Catalog,
Version 2 (Washington, DC, US Government Printing Office)

\bibitem {Lynch00} Lynch, D.K., Rudy, R.J., Mazuk, S. and Puetter, R.C. 2000, \apj,
 541, 791

\bibitem {Papad01} Papadopoulos, P., Ivison, R., Carilli, C. and Lewis, G. 2002,
Nature, 409, 59


\bibitem {Phinney89} Phinney, E.S. 1989, in {\bf Theory of Accretion Disks}, Meyer, F.,
Duschl, W.J., Frank, J. and Meyer-Hofmeister, E. eds., NATO ASI Series, 290, 457

\bibitem {Pier92} Pier, E.A. and Krolik, J.H. 1992, \apjl, 399, L23

\bibitem {Pier93} Pier, E.A. and Krolik, J.H. 1993, \apj, 418, 673


\bibitem {Sanders89} Sanders, D.B., Phinney, E.S., Neugebauer, G., Soifer, B.T. and 
Matthews, K. 1989, \apj, 347, 29

\bibitem {Salpeter74} Salpeter, E.E. 1974, \apj, 193, 579

\bibitem {Tran01} Tran, Q.D. et al. \apj, 552, 527

\bibitem {Wilkes01} Wilkes, B.J. 2001, New Astronomy Reviews, 45, 641

\end{thebibliography}
\end{document}